\documentclass[twocolumn,showpacs,superscriptaddress,preprintnumbers,amsmath,amssymb,prb]{revtex4-1}

\usepackage{graphicx}
\usepackage{dcolumn}
\usepackage{bm}
\usepackage{epsfig}
\usepackage{soul}
\usepackage{subfigure}
\usepackage{ulem}
\usepackage[usenames,dvipsnames]{color}

\begin{document}


\title{Realistic finite temperature simulations of magnetic systems using quantum statistics }

\author{Lars Bergqvist}
\affiliation{Department of Applied Physics, School of Engineering Sciences, KTH Royal Institute of Technology, Electrum 229, SE-16440 Kista, Sweden}
\affiliation{SeRC (Swedish e-Science Research Center), KTH Royal Institute of Technology, SE-10044 Stockholm, Sweden}
\author{Anders Bergman}
\affiliation{Maison de la Simulation, USR 3441, CEA-CNRS-INRIA-Universit\'{e} Paris-Sud-Universit\'{e} de Versailles, F-91191 Gif-sur-Yvette, France}
\affiliation{L\_Sim, INAC-MEM, CEA, F-38000 Grenoble, France}

\date{\today}

\begin{abstract}
We have performed realistic atomistic simulations at finite temperatures using Monte Carlo and atomistic spin dynamics simulations incorporating quantum (Bose-Einstein) statistics. The description is much improved at low temperatures compared to classical (Boltzmann) statistics normally used in these kind of simulations, while at higher temperatures the classical statistics are recovered. This corrected low-temperature description is reflected in both magnetization and the magnetic specific heat, the latter allowing for improved modeling of the magnetic contribution to free energies.
A central property in the method is the magnon density of states at finite temperatures and we have compared several different implementations for obtaining it.  The method has no restrictions regarding chemical and magnetic order of the considered materials. This is demonstrated by applying the method to elemental ferromagnetic systems, including Fe and Ni, as well as Fe-Co random alloys and the ferrimagnetic system GdFe$_3$ .
\end{abstract}

\maketitle


\section{Introduction}
Computer simulations of materials have developed to a very powerful alternative and complement to experimental methods. Electronic structure methods are now commonplace to predict novel ground state properties of existing and yet to be synthesized materials, sometimes in combination with screening of materials databases and applying artificial intelligence tools \cite{Ortiz2009,Jain2013,Pizzi2016}. These tools are only useful if the underlying methods are accurate enough to properly describe the materials. A common method to simulate finite temperature magnetic properties is to employ a two-step process where in the first step all material specific parameters are calculated within zero-temperature electronic structure theory to construct a simplified statistical physics model which is then solved in the second step by employing atomistic simulations. This methodology has been very successful of qualitatively predicting Curie temperatures and magnon spectra of transition metals, alloys and diluted magnetic semiconductors  \cite{Rosengaard1997,Pajda2001,Bergqvist2004,Ersogh2005,Lezaic2007,Kudrnovsky2008,Sato2010,Bergman2010,Bergqvist2013,Etz2015,ASDbook2017}. An overwhelming majority of the reported simulations of this kind have hitherto been based on classical Boltzmann statistics at finite temperature instead of the proper Bose-Einstein quantum statistics. In principle, quantum statistics are expected to enhance the critical temperature with the factor $(S+1)/S$, where $S$  is the spin quantum number. For large enough $S$ and at high enough temperatures, classical statistics has been shown to work reasonable well but at the same time, the limitations are well known. For instance, when using classical statistics the magnetization at low temperatures decreases too rapidly and the simulated low temperature magnetic specific heat is non-zero. Up until recently, there has been a lack of practical theories taking into account quantum statistics in atomistic simulations. 
An early effort was done by K\"{o}rmann \textit{et. al.} \cite{Kormann2011} where reducing a full atomistic description to a minimal nearest-neighbour model allows for a quantum Monte Carlo treatment of the reduced model. More recently, Woo. \textit{et. al.} \cite{Woo2015} demonstrated how it is possible to incorporate quantum Bose-Einstein statistics in existing atomistic simulation framework and obtained qualitatively good results for Fe in both low and high temperature region. In this work, we have extended and modified this latter methodology to more complex situations, such as random alloys and anti- and ferri- magnetic systems. We have also examined how the magnetic density of states, that is a central property of the model as discussed below, can be modelled and how the resulting quantum statistics is affected. Attempts to create a completely parameter-free self-consistent theory are discussed.

The paper is organized as follows: In Section~\ref{sec:Theory} we introduce the methodology and give the details of the calculations, in Section~\ref{sec:Results} we present our results and compare the different methodologies used for obtaining the magnon density of states. A summary and outlook of the study is finally provided in Section~\ref{sec:summary}.
\section{Theory} \label{sec:Theory}
\subsection{Control of temperature in atomistic simulations}
A statistical physics treatment using either Atomistic spin dynamics (ASD)\cite{UppASD,ASDbook2017} or Monte Carlo simulations (MC) uses the Hamiltonian in Eq.~(\ref{eq:H}) as a starting point. 
\begin{equation} \label{eq:H}
\mathcal{H} = -  \sum _{ij} J_{ij} \mathbf{m}_i \cdot \mathbf{m}_j ,
\end{equation}
where $J_{ij}$ denote the exchange interactions between two atomic magnetic moments $\mathbf{m}$ at site $i$ and $j$. We are using the sign convention in which ferromagnetic interactions are characterized by positive values of $J_{ij}$ and antiferromagnetic with negative sign.

Within ASD, the temporal evolution of the atomic moments $\mathbf{m}$ is governed by the Landau-Lifshitz-Gilbert (LLG) equations (see c.f. Ref.~\onlinecite{ASDbook2017} for a detailed account of the method). The inclusion of finite temperature effects is obtained through Langevin dynamics, by adding a stochastic Gaussian shaped field $\mathbf{b}_i(t)$ (white noise) to the effective magnetic field $\mathbf{B}_i = - \frac{\partial\mathcal{H}}{\partial\mathbf{m}_i}$, that acts on each atomic moment $i$. The stochastic, or thermal, field $\mathbf{b}_i(t)$ is characterized by 
\begin{eqnarray} 
\langle \mathbf{b}_i(t) \rangle & = & 0 \\  
\langle b_i^k (t) b_j^l (t') \rangle & = & 2 D_i \delta _{ij} \delta _{kl} \delta (t-t') \\
\nonumber
\end{eqnarray}
where $i$ and $j$ denote lattice sites, $k$ and $l$ the carteisian components and $D_i$ is the amplitude of the stochastic field given by
\begin{equation}
D_i = \dfrac{\alpha}{(1+\alpha ^2)} \dfrac{\eta}{\mu_B m_i},
\end{equation}
where $\alpha$ is the Gilbert damping parameter. Using the Fokker-Plank equation and searching for stationary solution, the amplitude of $D$ can be related to the temperature $T$ through the scaling factor $\eta$\cite{ASDbook2017}. 

If equilibrium properties are desired, Monte Carlo (MC) simulations based on the Hamiltonian, Eq.~(\ref{eq:H}) are normally more efficient than ASD simulations. The quantum corrections as outlined above can easily be included in Monte Carlo as well, and for simplicity we restrict the discussion here to the Metropolis algorithm \cite{Metropolis1953}. These temperature considerations are fully transferable to other importance sampling MC methods, including the heat-bath algorithm \cite{Miyatake1986}. The crucial step for advancing in time is the transition probability $W$ between two states $s$ and $s'$ in the Markov chain. In the Metropolis algorithm, the following form of $W$ is used
\begin{equation}
W(s \rightarrow s') =  \left\{ \begin{array}{ll} \mathrm{exp}(- \frac{\Delta E}{\eta})  & \textrm{if} \  \Delta E > 0 \\  1 & \textrm{otherwise} \end{array} \right. 
\end{equation}

Thus the temperature dependence in both ASD and MC simulations is governed by the scaling factor $\eta$.
\subsection{Quantum statistics}
In classical statistics, i.e. using Boltzmann distribution, the scaling factor $\eta$, is given by
\begin{equation}  \label{eq:eta}
\eta^c= k_\mathrm{B} T,
\end{equation}
where the superscript $c$ denotes classical for clarity. When transitioning from classical to quantum statistics, it is then this temperature dependent $\eta$ that is to be modified. Using quantum statistics with the Bose-Einstein distribution, Woo {\it et. al} \cite{Woo2015} showed that the relation in Eq.~(\ref{eq:eta}) is modified to the more complex form
\begin{equation} \label{eq:etaq}
\eta^q(T)= \int _0 ^\infty \frac{E}{\mathrm{exp}(E/k_\mathrm{B}T)-1} g(E,T) \mathrm{d}E,
\end{equation}
where $g(E,T)$ is the magnon density of states (mDOS) at temperature $T$ and $E$ the energy of a magnon at a wave-vector $\mathbf{q}$. Here, the superscript $q$ denotes quantum for clarity.

An alternative to the proper Bose-Einstein expression in Eq.~(\ref{eq:etaq}), is to perform the temperature rescaling based on empirical data, such as the experimental magnetization curve, which was proposed by Evans {\it et. al} \cite{Evans2015} using a simplistic model including free parameters. The big advantage with the present method is, depending on how the mDOS in Eq.~(\ref{eq:etaq}) is modelled, the number of free parameters can be reduced, or even eliminated.
\subsection{Magnon density of states}
The most important quantity entering the expression for the rescaled field arising from the quantum fluctuations controlling the temperature in Eq.~(\ref{eq:etaq}) is the temperature dependent magnon density of states (mDOS) $g(E,T)$ and in this section we discuss two different methods for obtaining it. 
\subsubsection{Adiabatic magnon spectra} \label{sec:ams}
The adiabatic magnon spectra (AMS)\cite{Halilov1998} is directly connected to the real-space exchange interactions as appear in Eq.~(\ref{eq:H}) through Fourier transformation of the Hamiltonian. A detailed account on how the procedure is performed is found in Appendix~\ref{app:ams} and the result is a set of eigenvalues $E(\mathbf{q}_i)$ depending on the wave-vector $\mathbf{q_i}$. By integrating the AMS over all wave-vectors, the resulting mDOS $g(E)$ is obtained.  In practice, in this work we convolute the mDOS with a Gaussian to ensure numerical stability, and then normalize it so that $\int_0^{\infty}g(E)dE=1$. 

The adiabatic mDOS is temperature independent and valid only at the ground state, i.e T=0 K. In order to create a temperature dependent mDOS  from the AMS, we apply the quasiharmonic approximation (QHA) as was outlined in Ref.~[\onlinecite{Woo2015}]. If $E_c(0)$ is the highest energy that give contribution to the mDOS, which is also put as upper limit in the integration in Eq.~(\ref{eq:etaq}), then at finite temperature ($T<T_c)$ the cutoff energy becomes rescaled as
\begin{equation} \label{eq:betacut}
E_c(T)=E_c(0) \cdot (1-\frac{T}{T_c})^\beta,
\end{equation}
where $T_c$ is the critical temperature and $\beta \approx 0.365$ the critical exponent associated with the magnetization in the 3D Heisenberg model. As pointed out in Ref.~[\onlinecite{Woo2015}], this particular model assumes that above $T_c$ the cutoff frequency is zero and thus the quantum statistics are sharply reverted back to classical statistics.
\subsubsection{Dynamical structure factor} \label{sec:SQW}
%
The magnon dispersion can also be obtained by simulating the dynamical structure factor $S(\mathbf{q},E)$ \cite{Bergman2010,Bergqvist2013,Etz2015} and identifying the peak values of $S(\mathbf{q},E)$ for each wave-vector $\mathbf{q}$. In contrast to the adiabatic treatment, temperature effects from the Gilbert damping processes are included so that a finite temperature description of the magnon dispersion is obtained. It is however worth noting that the present implementation assumes a fixed magnitude of the magnetic moments and does thus not include longitudinal fluctuations which give rise to Stoner excitations and an additional damping mechanism for magnons (Landau damping). The simulated finite temperature $g(E)$ is obtained from integrating $S(\mathbf{q},E)$ over all sampled wave-vectors. In order to minimize the numerical noise in the simulated magnon spectra and resulting mDOS, we perform a sampling and post-processing protocol as described in Appendix.~\ref{app:sqw}. 
\subsection{Details of calculations}
All first-principles calculations in this study was performed using a multiple-scattering (Korringa-Kohn-Rostoker, KKR) implementation of the density functional theory (DFT) in the local spin density approximation (LDA) as implemented in the SPR-KKR software \cite{SPRKKR}. The calculations were performed scalar relativistically using full potential and employing a basis set consisting of $spdf$-orbitals. The coherent potential approximation (CPA) was employed for treating the disordered FeCo alloy. The magnetic exchange interactions that appear in Eq.~(\ref{eq:H}) were obtained from the magnetic force theorem using the LKAG formalism \cite{Lichtenstein1987} in the low temperature magnetic reference state. The exchange interactions are treated as temperature independent in this study, however this approximation can be lifted using either noncollinear reference states\cite{Szilva2013} or using an extended spin model\cite{Pan2017}. The atomistic simulations, either the Monte Carlo or atomistic spin dynamics, were performed using the UppASD software \cite{UppASD,ASDbook2017}.

\section{Results} \label{sec:Results}
\subsection{Quasi-harmonic approximation (QHA)} \label{sec:qha}
\subsubsection{Simple cubic model with nearest neighbour interactions} \label{sec:NNsc}
\begin{figure}[htb]
\includegraphics[width=9cm]{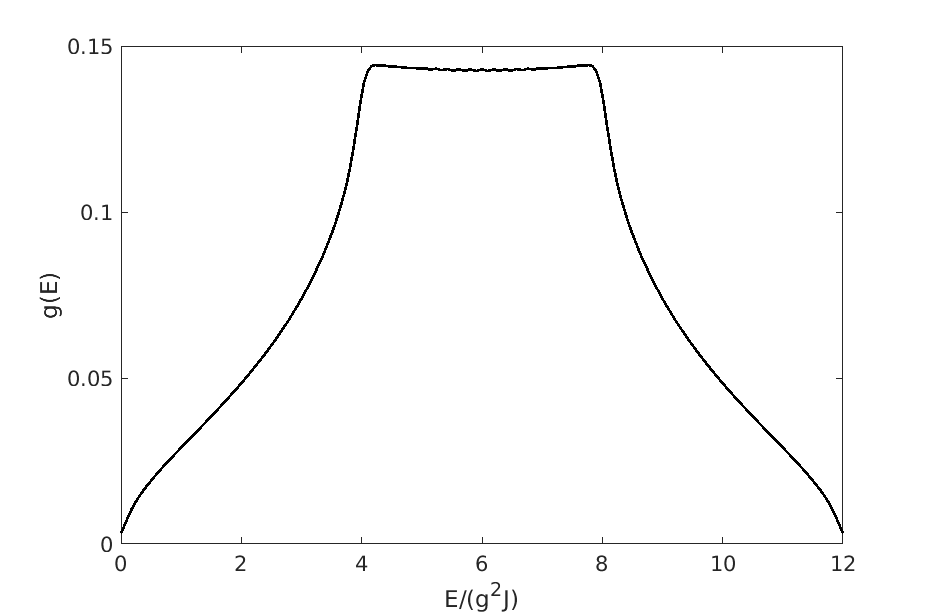}
\caption{Magnon density of states from AMS for bulk simple cubic ferromagnet with nearest neighbour interactions. }
\label{fig:mdos3dsc}
\end{figure}

As a first application of the method, we study a model system consisting of moments distributed on a bulk lattice in simple cubic geometry interacting with only nearest neighbours. It is a very well characterized system that has been served as a benchmark for many Monte Carlo simulations in the past. Although no analytical result exist for its Curie temperature, it has been calculated to very high precision \cite{Peczak1991}  to $k_BT_c/J \approx 1.4429$, which we also obtain (within error bars less than 0.005\%) using our software. The magnon density of states obtained from AMS is displayed in Fig.~\ref{fig:mdos3dsc}.

\begin{figure}[htb]
\includegraphics[width=9cm]{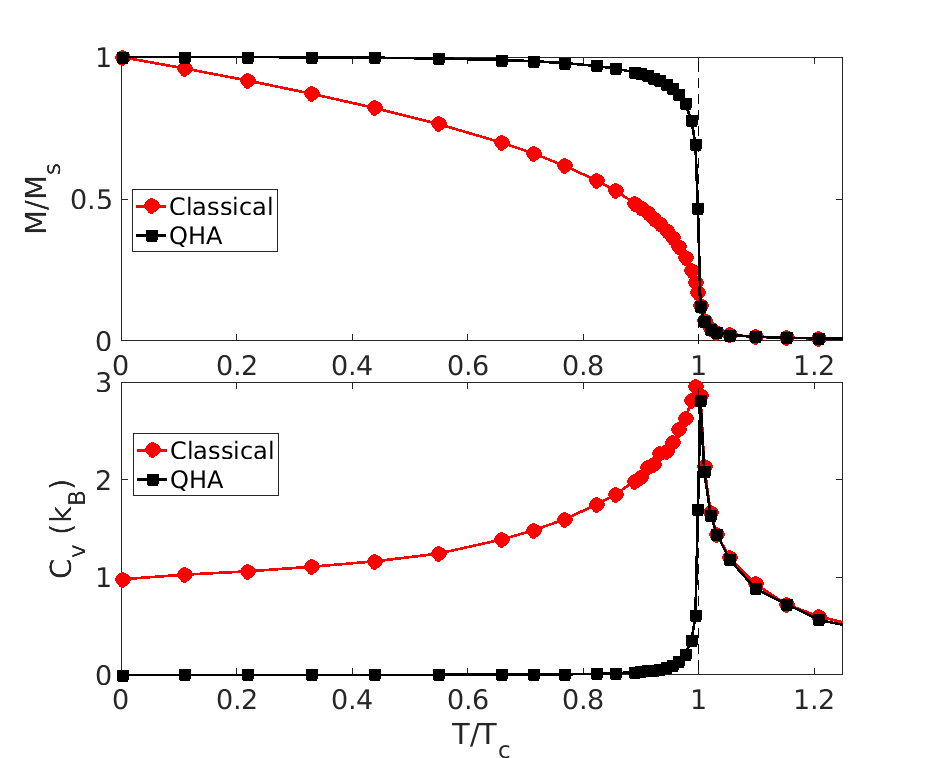}
\caption{Monte Carlo results of bulk ferromagnet with nearest neighbour interactions in simple cubic lattice. Red denotes classical statistics and black quantum statistics using QHA for the magnon density of states. (Upper) Reduced magnetization and (Lower) specific heat  as function of reduced temperature $T/T_c$   }
\label{fig:3dsc}
\end{figure}

In combination with the QHA in order to obtain the finite temperature mDOS, we show in Fig.~\ref{fig:3dsc} the results of the temperature dependence of the magnetization and the specific heat and compare those with the classical results. At low temperatures, it is well known that a classical description breaks down and the magnetization is decreasing too rapidly (linearly) with temperature and does not follow the Bloch $T^{3/2}$ law that is expected from linear spin wave theory analysis. Perhaps even more severe is the fact that the specific heat does not go to zero but to a constant value of 1 $k_B$ since each degree of freedom contributes with 1/2 $k_B$ and there are two transversal components. It can be seen as an analogy with the Dulong-Petit law for phonons where the classical value reaches a constant value of 3R, where R is the gas constant but where a quantum mechanical description in the form av either Einstein or Debye model obtains a vanishing specific heat when $T\to 0 K$. Similar, using the quantum statistics in the simulations, the magnon specific heat is approaching zero at low temperatures. In fact, the particular form of the mDOS for this system causes the fluctuations to be very small in a large temperature interval and only become significant in a narrow region close to $T_c$.
%
\subsubsection{FeCo alloys}
An important class of materials for technological applications are transition metal alloys such as Fe$_{1-x}$Co$_x$. Fe-Co alloys exhibit a large saturation magnetization and Curie temperature, larger than the values for Fe or Co separately and are commonly used in read head devices in magnetic storage applications. In addition, around the composition of Fe$_{0.75}$Co$_{0.25}$, the alloy has a measured Gilbert damping which is unusually low for a metallic system\cite{Schoen2016}.  A great advantage with the present method is the applicability to not only elemental magnetic systems but also magnetic alloys and compounds which we here demonstrate by simulating  finite temperature properties of Fe$_{0.5}$Co$_{0.5}$.
\begin{figure}[htb]
\includegraphics[width=9cm]{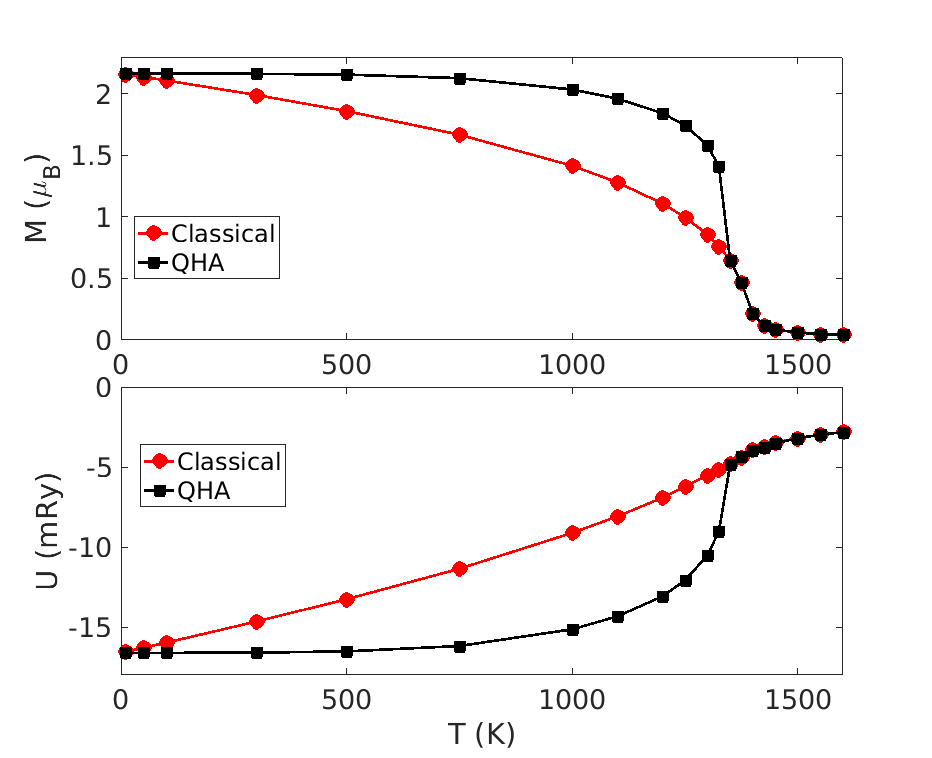}
\caption{Monte Carlo simulations of  Fe$_{0.5}$Co$_{0.5}$ random alloy in bcc structure using classical (red) and quantum (black) statistics and QHA. a) Magnetization and b) internal energy  as function of the temperature.}
\label{fig:FeCo50}
\end{figure}

 Fe-Co around 50-50 composition can either be synthesized in an ordered magnetic compound forming CsCl (B2) crystal structure or as a completely random alloy in the bcc crystal structure. We have run calculations on both these structures and the results were found very similar between the two. Hence, we decide to present only the results of the random alloy. Calculation of the mDOS from AMS of a random alloy is slightly more involved than for an elemental material due to the random arrangement of the atoms.  We therefore calculated the mDOS for 10 000 different disorder configurations and afterwards took the average of all those configuration in order to get the final averaged mDOS. 

The results from Monte Carlo simulation are displayed in Fig.~\ref{fig:FeCo50}. The shape of the magnetization is similar to that of elemental Fe, Section~\ref{sec:smdos} with the key difference however that calculated $T_c$ is higher, around 1395 K which is in good agreement with experiments\cite{Kawahara2003}. The quantization effects arising from using quantum statistics and QHA are also very evident when inspecting the internal energy of the system. At low temperatures, classical statistics overestimate magnon excitations causing a rapid increase of the internal energy in contrast to the internal energy from using quantum statistics that are much less affected. In the critical region around $T_c$, the quantization effects drastically reduces and vanish at $T_c$. 
\subsubsection{GdFe$_3$}
%
A phenomenological correction using renormalized temperatures to the magnetization shape for ferromagnetic materials were used in an earlier study\cite{Evans2015} but it has the shortcoming that it will not work for other magnetic orderings like antiferromagnets and ferrimagnets. In comparison, a great advantage with the present methodology it does not have this restriction which we will here demonstrate for the ferrimagnetic system GdFe$_3$. A ferrimagnetic is characterized by two sublattices with antiferromagnetic coupling but not fully compensating each other such as at low temperatures there is a finite (small) magnetization in the sample. GdFe$_3$ is the prototype system for so called all-optical switching\cite{Stanciu2007,Chimata2015} in which a laser pulse is used as a stimulus for magnetic switching where both the magnetic sublattices obtain a moment orientation opposite to the initial state after the pulse.  Crystalline GdFe$_3$, as treated here, has a trigonal unit cell (spacegroup R-3mH,166) with 3 different Fe sites and 2 different Gd sites. We employed the experimental crystal structure and volume in our calculations using the LDA+U approximation for the Gd $f$-states with a value of the Hubbard parameter $U$ set to 6.7 eV and Hunds exchange $J$ set to 0.7 eV. Experimentally, the critical and compensation temperatures of GdFe$_3$ are 721 K and 618 K, respectively \cite{GdFe3EXP}.
\begin{figure}[htb]
\includegraphics[width=9cm]{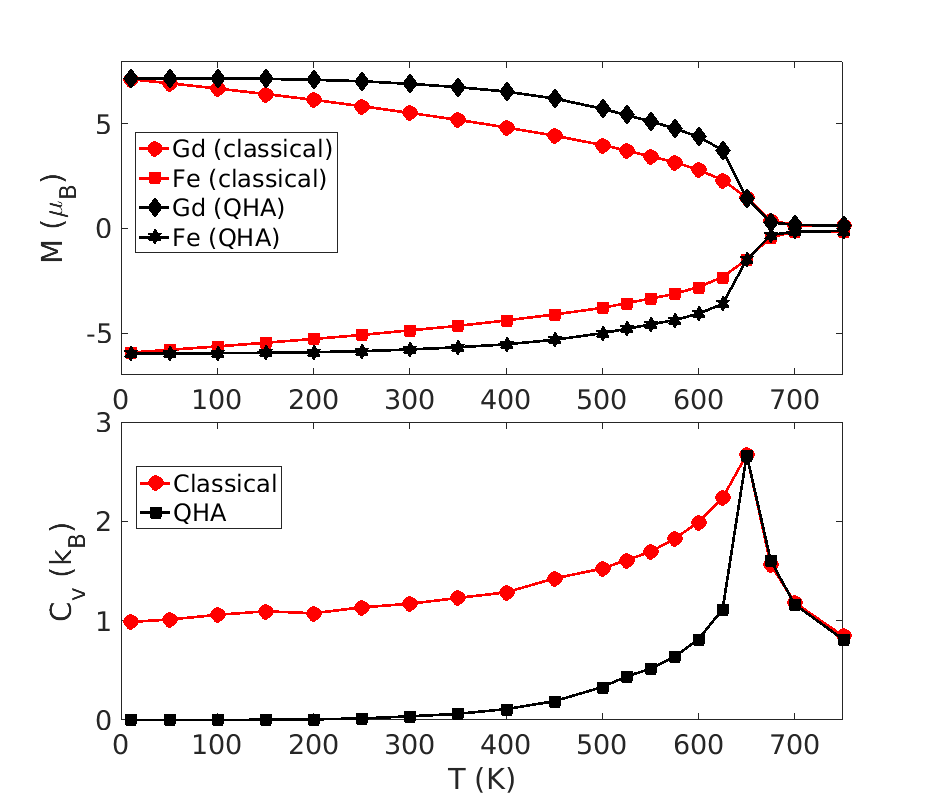}
\caption{(Upper) Sublattice magnetization per formula unit of GdFe$_3$ from Monte Carlo simulations. Red denoted classical statistics and black quantum statistics using QHA. (Lower) Specific heat as function of temperature. }
\label{fig:GdFe3}
\end{figure}
In the upper panel of Fig.~\ref{fig:GdFe3}, calculated Gd and Fe sublattice magnetization from Monte Carlo simulations are displayed. Each of the sublattice magnetization is behaving similar to the ferromagnetic systems in the sense that the magnetization is decreasing too rapidly with temperature when classical statistics are employed. This is corrected when using quantum statistics and applying the QHA to the mDOS which result in a much more stable magnetization with respect to temperature. Experimentally, the rare-earth (Gd) moment has a more rapid decrease of its sublattice magnetization than the Fe, causing a zero total moment at a temperature lower than $T_c$. This is the compensation temperature of a ferrimagnet. Taking the difference between our calculated sublattice magnetization, we do not obtain any compensation point, suggesting that an even more elaborate approach is needed to capture this behaviour, in particular additional longitudinal spin fluctuations of the moment. Nevertheless, we obtain a $T_c$ of around 650 K, in reasonable close agreement with experiments and keeping in mind that the calculated value is slightly dependent on chosen values of $U$ and $J$. The value of $T_c$ is even more pronounced when inspecting the peak of the specific heat as displayed in the lower panel of Fig.~\ref{fig:GdFe3}. Similar to ferromagnets, using quantum statistics recover the correct behaviour at low temperatures with vanishing specific heat and in turn the entropy (not shown).
\subsection{Simulated mDOS} \label{sec:smdos}
In order to study the difference between using QHA on the adiabatic magnon spectrum compared to using a simulated mDOS we here consider the finite temperature properties of bcc Fe. Fe in the body centered cubic (bcc) structure is the prototypical ferromagnet and often serves as a benchmark for testing and validating any new features in atomistic simulations. Experimentally, Fe has a Curie temperature $T_c$ of 1043 K and the magnetization curve as function of temperature is well known from experiments. The size of the magnetic moments in Fe is relatively robust with respect to rotations and temperature and therefore rather well described by an Heisenberg Hamiltonian. By calculating exchange parameters $J_{ij}$ up to three lattice constants from the low temperature ferromagnetic configuration and subsequent Monte Carlo simulations, we obtain $T_c$ of around 930 K, in agreement with previous studies \cite{Rosengaard1997,Pajda2001,Ruban2004}. Improvements to $T_c$ can be obtained by including electron entropy effects and/or include longitudinal spin fluctuations \cite{Ruban2007,Pan2017}, but that was not considered in this study.
\begin{figure}[htb]
\includegraphics[width=9cm]{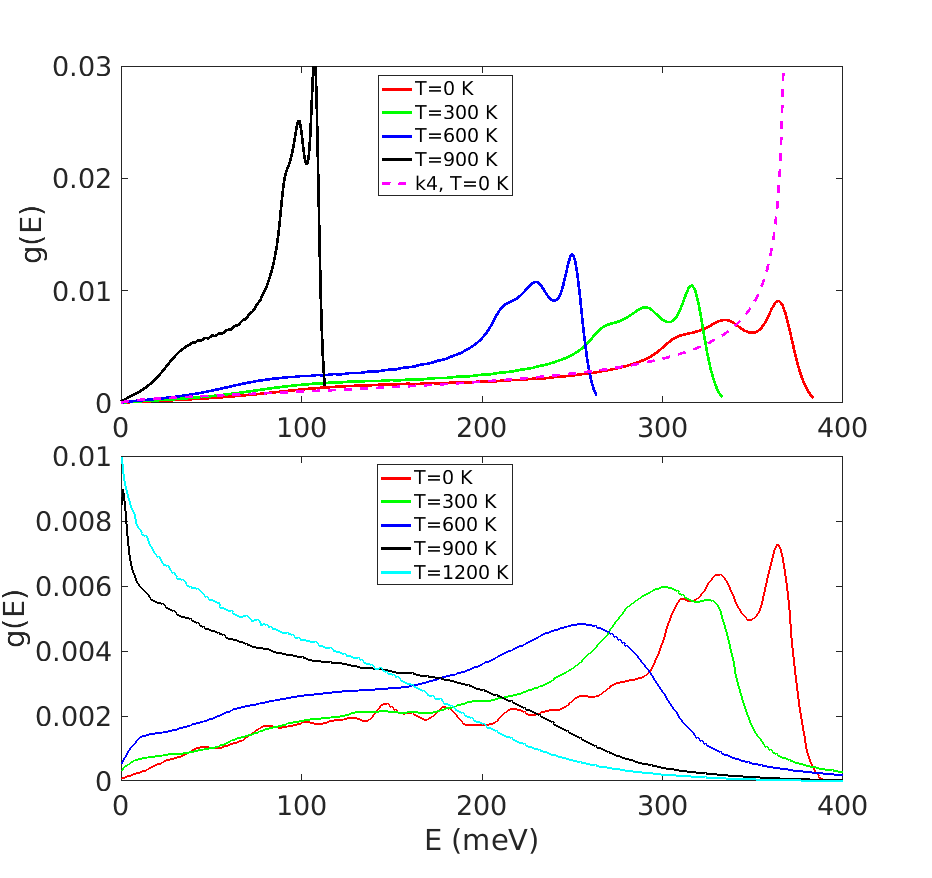}
\caption{ Temperature dependent magnon density of states of bcc-Fe using quasiharmonic approximation (upper panel) and dynamical structure factor (lower panel). "k4" denotes the simplified model mDOS from Ref.[\onlinecite{Woo2015}].}
\label{fig:bccFeMDOS}
\end{figure}

As emphasized by Eq.~(\ref{eq:etaq}), the mDOS is a central quality to extracting the quantum statistical description of thermal excitations. How the temperature dependence of the mDOS is modelled is thus directly determining the resulting statistics. Results for the finite temperature mDOS from AMS using the QHA, i.e. the same methodology as was used for the systems in Section~\ref{sec:qha}, are displayed in the upper panel of Fig.~\ref{fig:bccFeMDOS}. Qualitatively the spectral weight is shifting towards lower frequencies with increasing temperatures and vanishes at $T_c$, where the distribution become continuous i.e. classical statistics apply. The  sharp transition from quantum to classical statistics at $T=T_c$ comes directly from the QHA model through the rescaling in Eq.~(\ref{eq:betacut}). This can be motivated by the fact that the spin stiffness vanishes along with the magnetization at $T_c$. However, there are experimental and theoretical indications of magnons even above $T_c$\cite{Qin2017,Tao2005} and thus the validity of the QHA cut-off temperature can be questioned. The scaling in Eq.~(\ref{eq:betacut}) does thus infer that $T_c$ is a free parameter in the model. In some aspects that can be a warranted feature, for example when using the experimental $T_c$ for an empirical description of the magnetization behaviour. Since the $T_c$ that is used as input for the QHA rescaling can be determined from MC simulations, using calculated exchange interactions, it can be argued that the $T_c$ is not a free parameter in that case. 
Another step towards a parameter-free description is to use a simulated mDOS instead of a model one. To obtain the simulated finite temperature magnon densities of states, we perform atomistic spin dynamics simulation to sample the excitation spectra represented by the dynamical structure factor $S(\mathbf{q},E)$, as described in Sec.~\ref{sec:ams} and Appendix~\ref{app:sqw}, which is then integrated to obtain $g(E,T)$. We will in the following refer to this combination for obtaining the simulated mDOS as the QHB method. The resulting simulated mDOS curves are shown in the lower panel of Fig.~\ref{fig:bccFeMDOS} . At low temperatures well below $T_c$, the results from QHA and the simulated QHB results agree very well as expected. For elevated temperatures the trend of the QHB mDOS follows the trend from the QHA based data, with spectral weight transfer to lower frequencies with increasing temperatures. However, at high temperatures close to $T_c$, the difference of mDOS from QHA and QHB becomes evident. Instead of a vanishing mDOS as enforced by QHA, the QHB simulations yield an mDOS that fits rather well to a classical Boltzmann distribution. 

\begin{figure}[htb]
\includegraphics[width=9cm]{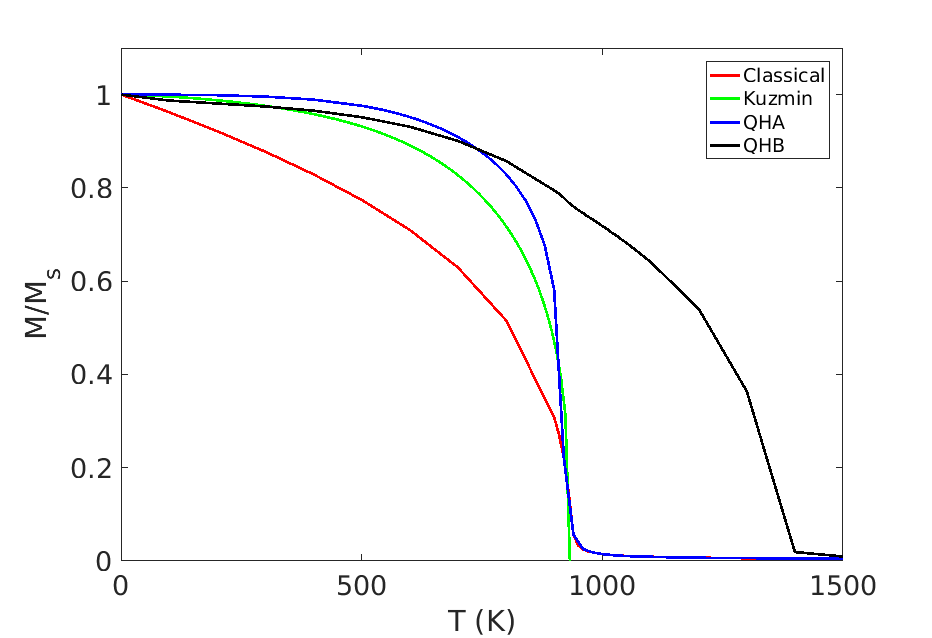}
\caption{Temperature dependent magnetization of bcc-Fe from Monte Carlo simulations in classical and quantum statistics using magnon density of states from either quasi-harmonic approximation (QHA) or dynamical structure factor (QHB), as well as comparison with model shape of Kuz'min from Ref.~[\onlinecite{Kuzmin2005}].}
\label{fig:bccFeM}
\end{figure}

%
The simulated mDOS is then used as input for the determination of the temperature rescaling factor $\eta^q$ for which MC simulations give the resulting temperature dependent magnetization as displayed in Fig.~\ref{fig:bccFeM}. For comparison with experiments, an empirical magnetization curve according to Kuz'min\cite{Kuzmin2005} is also shown in Fig.~\ref{fig:bccFeM}. The Kuz'min curve follows an analytic expression for the shape of the magnetization that reproduces the experimental curve by matching the low temperature Bloch $T^{3/2}$ law with the high temperature critical behaviour resulting in the expression
\begin{equation} \label{eq:Kuzmin}
m(t)=(1-st^{3/2}-(1-s)t^p)^{1/3},
\end{equation}
where $m(t)=M(T)/M(0)$, $t=T/T_c$, $s$ and $p$ parameters with values 0.35 and 4.0, respectively.

Using classical statistics, we obtain the usual shortcomings that the magnetization drops too rapidly at low temperatures and has a linear temperature dependence instead of the expected Bloch $T^{3/2}$ behaviour. However, using quantum statistics and QHA, apart from small deviations close to $T_c$ which arises from finite size effects from the simulations, we obtain a similar shape as the empirical curve from Eq.~(\ref{eq:Kuzmin}) (using our calculated $T_c$ and M(0)). This agreement is obtained regardless if the underlying mDOS is taken from AMS or the simplified quartic model mDOS from Ref.[\onlinecite{Woo2015}] (not shown). From the observation that the magnetization curve is very similar for the simplified quartic model and for the AMS mDOS it can be noted that the quartic model is an acceptable approximation under QHA since minor differences in the high energy region of the mDOS does not seem crucial for determining the M vs T behaviour. Why the shape of the high energy mDOS is less important can readily be understood by the exponential prefactor to the mDOS in the determination of the scaling factor $\eta^q$ as expressed in Eq.~(\ref{eq:etaq}).

The simulated temperature dependent mDOS within the QHB method, is much more general in the sense that it does not require any additional parameter and should thus yield a more realistic description of the magnon properties at finite temperatures. At lower temperatures, the magnetization is similar to the QHA treatment but at elevated temperatures two pronounced differences can be noticed. The curvature of the QHB curve is less pronounced, resulting in a flatter shape compared to both the Kuz'min and the QHA curves. More important, the $T_c$ obtained from the QHB curve differs as well. That means that when $T_c$ is not given as a parameter to the model, as in the QHA case, the use of quantum instead of classical statistics also affects the critical temperature of the simulated system. 

\begin{figure}[htb]
\includegraphics[width=9cm]{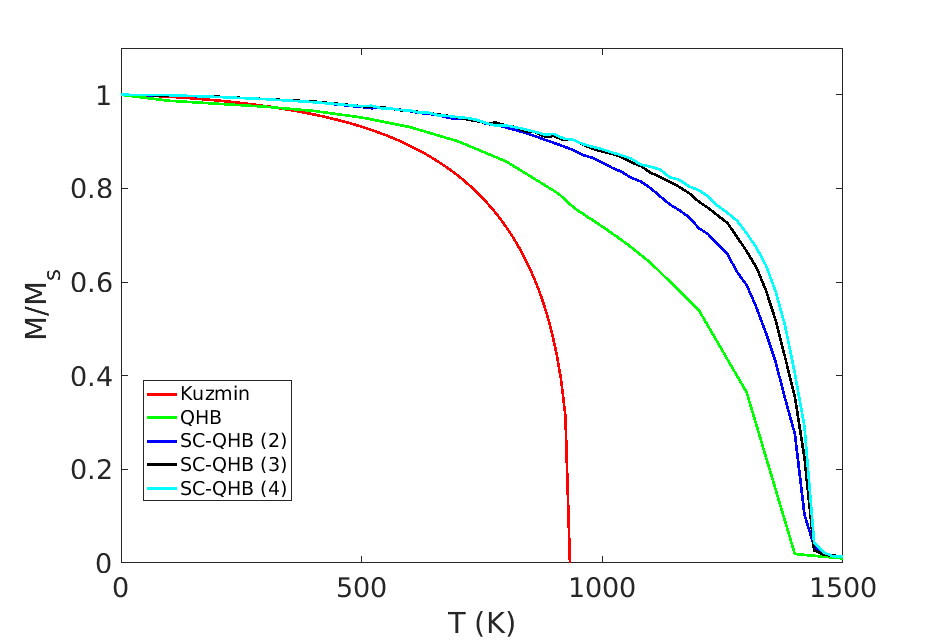}
\caption{Temperature dependent magnetization of bcc-Fe from Monte Carlo simulations using iterative self-consistent implementation of the QHB method with simulated mDOS, as well as comparison with model shape of Kuz'min from Ref.~[\onlinecite{Kuzmin2005}]. }
\label{fig:bccFeSCF}
\end{figure}

In the results above, the QHB mDOS was obtained using classical statistics and then applied to get a quantum statistical description of the following MC simulations. Since the use of quantum statistics affects the thermodynamics of the simulations, good arguments can be made that the QHB mDOS should also be obtained from quantum statistical simulations. This can be achieved in a self-consistent way by performing first a classical simulation, to get the first mDOS, and then successive QHB simulations, always using the mDOS obtained from the previous run, until the M vs T curves are converged. How this iterative approach performs can be seen in Fig.~\ref{fig:bccFeSCF} where it is found that a self-consistent result is obtained after 3-4 QHB simulations. Comparing the self-consistent results in Fig.~\ref{fig:bccFeSCF} with the single-shot output in Fig.~\ref{fig:bccFeM} it can be seen that $T_c$ is slightly increased further and also that the shape of the curve becomes more similar to the Kuz'min and QHA results.

In the case of bcc Fe considered here, the obtained increase in $T_c$ when using the self-consistent QHB mDOS could be seen as unfortunate since the classical simulations give a better agreement with the experimental $T_c$ compared to the quantum statistical QHB simulations. However, the increase in $T_c$ is consistent with picture that the critical temperature would have an enhancement factor of $\frac{(S+1)}{S}$, where $S$ is the spin quantum number, in a fully quantum mechanical treatment\cite{Lichtenstein1987} (the enhancement factor is equal to 1 in classical  statistics). Following the similar arguments as in Ref.~[\onlinecite{Lichtenstein1987}], the good agreement in classical statistics may be due to mutual error cancellation of various effects, for instance the use of temperature independent exchange interactions. By correcting the statistics, perhaps it is also needed to combine with more sophisticated methods to compensate, such as temperature dependent exchange interactions.

\begin{figure}[htb]
\includegraphics[width=9cm]{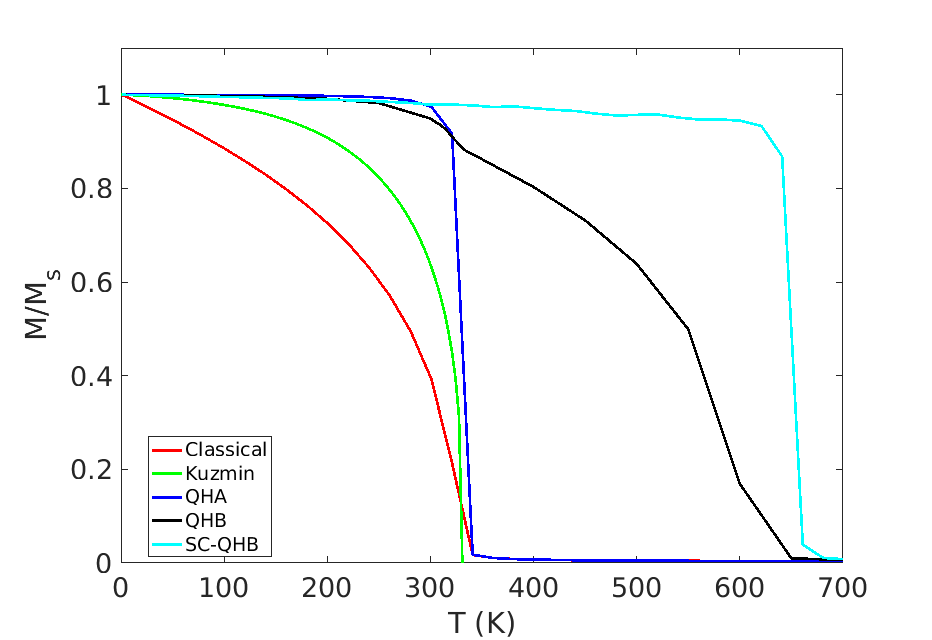}
\caption{Temperature dependent magnetization of fcc-Ni from Monte Carlo simulations using classical and quantum statistics in comparison with the empirical model shape of Kuz'min from Ref.~[\onlinecite{Kuzmin2005}]. The quantum statistical simulations are performed in either QHA, QHB or iterative self-consistent QHB (SC-QHB) with the results from the final converged iteration displayed. }
\label{fig:fccNiSCF}
\end{figure}

As an example where an enhanced $T_c$ from using quantum statistics is improving the theoretical results with respect to comparison with experiments, we show results for QHA and QHB simulations for fcc Ni in Fig.~\ref{fig:fccNiSCF}. Here we see that the increase in $T_c$ from 330 K in the classical case to 661 K, as obtained by self-consistent QHB simulations, is indeed an improvement when compared to the experimental value of $T_c~628-631 K$\cite{Kuzmin2005}. Since Ni has a smaller moment and thus lower spin quantum number than Fe, quantum effects are indeed expected to be larger in Ni. However, it is also seen that the steep shape of the simulated M vs T curve is strongly exaggerated compared to the experimental curve. We have not been able to figure out why the simulated M vs T curve takes this form but there are several potential reasons for it. First of all, it is well known that the exchange interactions in Ni are strongly temperature dependent and differs considerably depending on reference state. Related, the finite temperature description of Ni from a pure Heisenberg model is very poor, in particular it is needed to also include longitudinal spin fluctuations that are responsible for stabilizing a finite magnetic moment at high temperatures. Nevertheless, in many cases it is mostly interesting to obtain reliable values of $T_c$  and in those cases, the present method may provide that without too much complications.

\subsection{Thermodynamic properties}

\begin{figure}[htb]
\includegraphics[width=9cm]{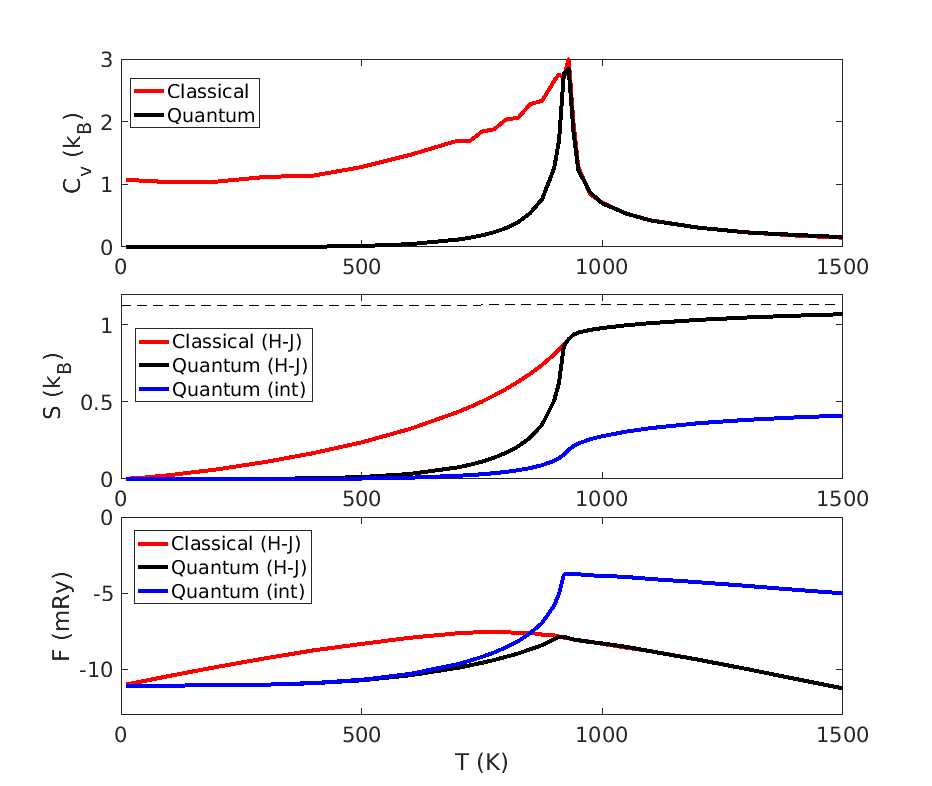}
\caption{ Thermodynamic properties of bcc-Fe from Monte Carlo simulations, (upper) specific heat $C_v$, (middle) magnetic entropy $S$ and (lower) free energy $F$. "H-J" denotes the Heine-Joynt model and "int" direct thermodynamic integration, see text.}
\label{fig:Fethermo}
\end{figure}

We now turn our attention towards how a quantum statistical description can improve the calculation of the free energy which is crucial in order to correctly determine phase stabilities. Returning to bcc Fe, relevant thermodynamic properties are gathered in Fig.~\ref{fig:Fethermo}. The specific heat $C_v$, that measures the amount of fluctuations in the system, has the familiar shape with a peak at $T_c$ and large difference between classical and quantum treatment where the latter is required in order to obtain the physical result $C_v \to 0$ when $T\to 0 $K. The magnetic entropy can be obtained from Monte Carlo simulations in two alternative ways, either by using a model or by direct thermodynamic integration. In the paramagnetic limit, the (transversal) magnetic entropy $S/k_\mathrm{B}=\mathrm{log}(M+1)$, where $M$ is the size of each atomic moment. Without accounting for longitudinal spin fluctuations, using the value $M=2.2\mu_B$, the high temperature magnetic entropy has value of 1.16$k_\mathrm{B}$, indicated by a dashed line in the middle panel of Fig.~\ref{fig:Fethermo}. In the opposite end, from the third law of thermodynamics, it is required that the entropy is zero when approaching zero temperature. In order to have a simplified model between these two limits, Heine and Joynt\cite{Heine1988} formulated a model interpolating between these two limits, later employed by Grimvall\cite{Grimvall1989} for fcc Fe. In this model, which we denote the Heine-Joynt model, the magnetic entropy has the following expression:

\begin{equation} \label{eq:HJ}
S/k_\mathrm{B}=t^3 \mathrm{log}(Mt^{-3}+1),
\end{equation}
where $t=2\phi/\pi$ and $\phi$ is the average angle between nearest neighbour magnetic moments. The angle $\phi$ and its temperature dependence is easily obtained in Monte Carlo simulations from the static correlation function and in turn the magnetic entropy using Eq.~(\ref{eq:HJ}). An alternative and more general method to obtain magnetic entropy is through thermodynamic integration

\begin{equation} \label{eq:Sint}
S(T)/k_\mathrm{B}=\int _0 ^T \frac{C_v}{T} \mathrm{d}T.
\end{equation}
From this expression, it is clear that classical statistics fail at low temperatures since at any non-zero temperature, the entropy will be finite. From our results, we obtain a lower entropy in the whole temperature interval from the thermodynamic integration compare to the Heine-Joynt model. It is however worth stressing that only the transversal fluctuations contribute to the entropy in these calculations, at elevated temperatures the longitudinal fluctuations of the moments\cite{Pan2017} have important contributions to the entropy. Once the magnetic entropy is obtained, it is trivial to obtain the (Helmholtz) free energy $F=U-TS$, where $U$ is the internal energy which is shown in the bottom panel of Fig.~\ref{fig:Fethermo}.

\section{Summary} \label{sec:summary}
A general improvement of finite temperature properties of magnetic systems from atomistic simulations are found using quantum statistics instead of classical statistics. The quantum effects enters through the temperature dependent magnon density of states. A simple but very effective method is the quasi harmonic approximation where the easy obtainable zero temperature mDOS is rescaled at finite temperatures and vanishing above the critical temperature in which the method is reverting back to classical description. The methodology has here been shown to work not only for ferromagnetic systems but also for anti- and ferri- magnetic magnetic systems as well as for disordered alloys. Perhaps the main shortcoming is the required knowledge of the critical temperature which separate the quantum and classical description such that the simulations must be performed in two steps, first to determine the critical temperature using classical statistics and secondly rerun the simulations using quantum statistics. To eliminate this shortcoming, we have constructed a more sophisticated approach where the temperature dependent magnon density of states is self-consistently obtained through magnetodynamical simulations. With this scheme it, the low-temperature behaviour is closely agreeing with the QHA method, where both methods have the important properties that the M vs T curve follows the Bloch $T^{3/2}$ and that the magnetic specific heat goes to zero at low temperatures. The latter property is crucial in order to be able to extract a good description of the magnetic contribution of the free energy. At larger temperatures, we notice that our suggested scheme gives an enhanched critical temperature $T_c$ compared to when using classical statistics. The method is already suitable to employ for atomistic simulations and free energy calculations and have the prospect of being improved further if longitudinal fluctuations and/or coupled spin-lattice contributions to the thermodynamics of the simulated systems are taken into account. Such efforts are ongoing.
\begin{acknowledgments}
The authors thank Danny Thonig for fruitful discussions. We acknowledge support from the VR (Swedish Research Council), SeRC (Swedish e-Science Research Centre), GGS (G\"oran Gustafssons Stiftelser), eSSENCE, 
and the CEA-Enhanced Eurotalents program, co-funded by FP7 Marie Sk\l{}odowska-Curie COFUND Programme (Grant Agreement n\textsuperscript{o} 600382). The computations were performed on resources provided by SNIC (Swedish National Infrastructure for computing) at NSC (National Supercomputer Centre) in Link\"oping, Sweden.
\end{acknowledgments}
%
%
\appendix
\section{Adiabatic magnon spectra and mDOS}  \label{app:ams}
Let $J^{\alpha\beta}(\mathbf{q})$ denote the Fourier transform of the exchange interaction between atom type $\alpha$ and $\beta$ with a wave-vector $\mathbf{q}$ lying in the Brillouin zone (BZ). $J^{\alpha\beta}(\mathbf{q})$ is calculated as
\begin{equation}
J^{\alpha\beta}(\mathbf{q})=\sum_{j\ne0} J_{0j}^{\alpha\beta} \mathrm{exp}(i \mathbf{q} \cdot \mathbf{R}_{0j}),
\end{equation}
where $\mathbf{R}_{0j}$ is a position vector connecting the two sites.  If there are $N$  atoms per unit cell, for each wave-vector $\mathbf{q}$, the magnon energies $\{E_1,E_2,...,E_N\}$ will then be given by the eigenvalues of a general $N \times N$ matrix $\Delta(\mathbf{q})$\cite{Halilov1998,Anderson1963} here expressed in block form
\begin{equation} \label{eq:AMS}
4  \left[ \begin{array}{cc}  \frac{1}{M_\alpha}   \sum_\gamma^N  \left(  J^{\alpha\gamma}(\mathbf{0})-J^{\alpha\alpha}(\mathbf{q}) \right) & 
-\frac{J^{\alpha\beta}(\mathbf{q})}{M_\alpha} \\
-\frac{J^{\alpha\beta}(\mathbf{q})^*} {M_\beta} &
\frac{1}{M_\beta} \sum_\gamma^N  \left(  J^{\gamma\beta}(\mathbf{0})-J^{\beta\beta}(\mathbf{q}) \right)
\end{array}
\right]
\end{equation}
From the magnon frequencies, the adiabatic mDOS $g(E)$ is obtained by summing up the energies as
\begin{equation}\label{eq:qsum}
g(E) = \sum _{q \in BZ} \sum_{i=1}^{N} \delta(E-E_i(q)).
\end{equation}
\section{Simulated magnon spectra and mDOS} \label{app:sqw}
By measuring the trajectories of the simulated atomic moments, the time and space correlation function is obtained
\begin{equation} 
\label{c_k}
C^{\alpha \beta}(\mathbf{r},t)=\frac{1}{N}\sum_{\substack{i,j ~ \text{where} \\\ ~\mathbf{r}_i-\mathbf{r}_j=\mathbf{r}}} \langle m_i^\alpha(t) m_j^\beta(0) \rangle -  \langle m_i^\alpha(t) \rangle \langle m_j^\beta(0) \rangle.
\end{equation}
The correlation function defined in Eqn.~(\ref{c_k}) can thus describe how the magnetic order evolves both in space and over time. The perhaps most valuable application of $C(\mathbf{r},t)$ is however obtained by a Fourier transform over space and time to give the dynamic structure factor 
\begin{equation} 
\label{sqw}
S^{\alpha \beta}(\mathbf{q},E)=\frac{1}{\sqrt{2\pi}N}\sum_{\mathbf{r}} e^{i\mathbf{q}\cdot\mathbf{r}} \int_{-\infty}^\infty e^{iE t}C^{\alpha \beta}(\mathbf{r},t) \,dt.
\end{equation}
Due to a combination of finite size effects, finite sampling time, temperature fluctuations and damping in the simulations, the raw output of the magnon density of states from simulations has associated noise that needs to be taken care of. For this purpose we employ a Hann window function for the time-domain transform and also apply a Gaussian convolution of the summed up mDOS combined with a normalization. In addition to the noise, there are also other numerical difficulties with the sampling of $S^{\alpha \beta}(\mathbf{q},E)$. While the correlation function is formulated in cartesian coordinates $\alpha,\beta$, the proper correlations stemming from the magnons in the system is only sampled in the spatial directions aligned perpendicularly to the magnetization. That means that in order to get a good mDOS from the dynamic structure factor, the perpendicular components of it must be identified, and more importantly, the longitudinal component should not be sampled. For that reason, we perform a rotation of the measured trajectories towards the quantization axis so that $S^{\parallel}(\mathbf{q},E)$ and $S^{\perp}(\mathbf{q},E)$ is sampled. At finite temperatures, there can however be a small drift of the magnetic system due to the temperature dependent stochastic fluctuations present in the simulations so defining a static quantization axis might prove difficult. As a result, the simulated mDOS might show spurious peaks for the low-energy part of the spectrum, in particular at $E=0$. While these peaks can be partially suppressed by Gaussian convolution, we have found that a better way to remove the spurious peaks is to update the quantization axis over time during the simulations. Physically, this can be motivated by the fact that the Goldstone mode of a ferromagnet should not carry any spectral weight in the mDOS. The quantization axis update is however performed slowly compared to the time-scale of the simulations in order to minimize the effect on the simulated spectra at $E>0$.

%

\end{document}